# Electromagnetic dynamical characteristics of a surface plasmon-polariton

A. Y. Bekshaev, L. V. Mikhaylovskaya

I.I. Mechnikov National University, Physics Research Institute,
Dvorianska 2, 65082, Odessa, Ukraine,
e-mail: bekshaev@onu.edu.ua

**Abstract**

We consider an electromagnetic field near the interface between two media with arbitrary real frequency-dependent permittivities and permeabilities, under conditions supporting the surface plasmon-polariton (SPP) propagation. The dispersion of the electric and magnetic properties is taken into account based on the recent approach for description of the spin and momentum of electromagnetic field in complex media [*Phys. Rev. Lett.* **119**, 073901 (2017); *New J. Phys.*, **19**, 123014 (2017)]. It involves the Minkowski momentum decomposition into spin and orbital parts with the dispersion-modified permittivities and permeabilities. Explicit expressions are derived for spatial densities of the energy, energy flow, spin and orbital momenta and angular momenta of the transverse-magnetic (TM) SPP field. The expressions are free from non-physical singularities; the only singular contribution describes a strictly localized surface part of the spin momentum that can be associated with the magnetization current in the conductive component of the SPP-supporting structure. On this ground, a phenomenological theory of the SPP-induced magnetization (predicted earlier based on the simplified microscopic approach) is outlined. Possible modifications and generalizations, including the transverse-electric (TE) SPP waves, are discussed.

OCIS codes: 240.6690 (Surface waves); 240.6680 (Surface plasmons); 240.5420 (Polaritons); 260.2160 (Energy transfer); 260.2110 (Electromagnetic optics); 230.3810 (Magneto-optic systems)

Keywords: surface plasmon-polariton; electromagnetic momentum; optical spin; dispersive media; spin-orbital decomposition; magnetization.

## 1. Introduction

During the past decades, a significant interest has been attracted to the study of localized optical fields associated with interfacial regions between nanostructured metals and dielectrics [1–6]. Especially, the running evanescent surface waves (surface plasmon-polaritons, SPP) are intensively investigated in connection with the optical nano-probing and precise optical manipulation. Additionally, the SPP fields pave new ways for the light wave-guiding, switching and controlling by sub-wavelength elements, which is crucial for further microminiaturizing of optical information devices and systems. Many applications are stimulated by the unique dynamical properties of the evanescent waves and SPPs, in particular, the transverse spin and momentum [7–9], the special spin-momentum locking [9,10], nonreciprocity and unidirectional propagation [9–12].

Earlier, the wide application of SPP-based techniques was restricted by the rather special requirements to materials that support efficient SPP generation and propagation [2–5]. However, with advent of a novel class of engineered composite materials, including the metamaterials, left-handed materials, sculptured films, etc. [4,13–16], almost any combination of the electric and magnetic parameters in the optical frequency range is becoming available [4,10,17–19], which offers additional and unexpected possibilities for the research and applications. As a rule, exclusive properties of new materials are accompanied by the strong dispersion (frequency dependence of the main electric and magnetic parameters), which is also a characteristic feature of the SPP waves. In this situation, the existing means for theoretical description of an SPP field, especially its dynamical characteristics (energy, energy flow, momentum, angular momentum and their derivatives) become insufficient and ambiguous. For example, existence of different definitions of the field momentum (Abraham or Minkowski paradigms) leads to contradictory and hardly interpretable results; the strong spatial inhomogeneity (interface between two media with radically different electromagnetic properties) and strong dispersion still reinforce the need in generalization and elaboration of adequate theoretical instruments for the SPP field characterization.

Recently, a new approach to description of the electromagnetic field dynamical characteristics in complex lossless media has been developed [20–24] based on the canonical (spin-orbital) decomposition [25–27] of the Minkowski momentum. Notably, its abilities in characterizing the "structured light in structured matter" has been convincingly demonstrated with examples of the SPP fields. However, because of the illustrative character, in [22–24] only the simplest SPP-supporting structure was considered which is formed by a non-magnetic ideal metal (described by the standard plasma model (SPM) [5]) and a vacuum. In this paper we intend to generalize this approach and to develop a meaningful description of the dynamical characteristics and associated SPP properties for a wide range of the electric and magnetic parameters of the contacting media.

The general model of the SPP-supporting structure is presented in Section 2 where the known expressions for the magnetic and electric fields in the transverse-magnetic (TM) surface mode are revisited; the partnering media are characterized by arbitrary phenomenological permittivities and permeabilities with arbitrary dependence on frequency. In Section 3 these expressions are used for calculating the spatial distributions of the field energy, energy flow (Poynting vector), spin and orbital (canonical) momenta [8,26,27], as well as the spin and orbital angular momenta. In contrast to other recent approaches [7–10,22–24], the consideration is consistently phenomenological with allowance for arbitrary dispersion of the conductive medium by means of the dispersion-modified dielectric and magnetic constants [22–24]. The results appear to be physically consistent, with no physically irrelevant singular terms that occur with other approaches (see, e.g., [7]). The only exclusion is the singular surface part of the spin momentum that is interpreted in Section 4 as a surface magnetization current. This allows us to develop a simple phenomenological theory of the SPP-induced magnetization predicted earlier [22,23] on the base of the microscopic model. In Section 5, the analytical results obtained are illustrated numerically by examples of the SPP fields formed near the interfaces between vacuum and silver or golden films. Finally, the main limitations of the presented approach, its possible generalizations and extension to the transverse-electric (TE) modes are briefly discussed in Section 6.

## 2. General description of the model

We consider a standard system supporting the SPP propagation [3–5] (Fig. 1a). Two homogeneous media are separated by the plane interface $x = 0$; medium 1 ($x > 0$) is dielectric with electric and magnetic constants $\varepsilon_1$ and $\mu_1$, medium 2 ($x < 0$) is conductive (a metal) characterized by $\varepsilon_2(\omega)$ and $\mu_2(\omega)$. The electromagnetic field is monochromatic, and instantaneous values of the electric and magnetic vectors are represented via the complex amplitudes: $\mathcal{E}(t) = \text{Re}\left(\mathbf{E}e^{-i\omega t}\right)$, $\mathcal{H}(t) = \text{Re}\left(\mathbf{H}e^{-i\omega t}\right)$. The frequency dependence of the electric and magnetic properties (optical dispersion) in the conductive part of the structure is crucial for the SPP physics, and it is taken into account; the medium 1 is supposed to be non-dispersive. For simplicity and in compliance with the approach of [19,20], we consider a lossless system, i.e. all the permittivities and permeabilities are real.

In each medium, the Maxwell equations should be satisfied [28,29]

$$\nabla \mathbf{H} = 0, \quad \mathbf{H} = \frac{1}{ik\mu}\nabla \times \mathbf{E},$$

$$\nabla \mathbf{E} = 0, \quad \mathbf{E} = -\frac{1}{ik\varepsilon}\nabla \times \mathbf{H} \qquad (1)$$

where $k = \omega/c$, $c$ is the speed of light in vacuum, and the Gaussian system of units is used. For the TM solutions of the Maxwell equations, the boundary conditions require that at $x = 0$ $E_{z1} = E_{z2}$, $H_{y1} = H_{y2}$ and $\varepsilon_1 E_{x1} = \varepsilon_2 E_{x2}$. As a result, the electric and magnetic vectors of the SPP field are obtained in the form [3–7]

$$\mathbf{E}_1 = \frac{A}{\varepsilon_1}\left(\mathbf{x} - i\frac{\kappa_1}{k_s}\mathbf{z}\right)\exp(ik_s z - \kappa_1 x), \quad \mathbf{H}_1 = \mathbf{y}A\frac{k}{k_s}\exp(ik_s z - \kappa_1 x), \quad (x > 0); \qquad (2)$$

$$\mathbf{E}_2 = \frac{A}{\varepsilon_2}\left(\mathbf{x} + i\frac{\kappa_2}{k_s}\mathbf{z}\right)\exp(ik_s z + \kappa_2 x), \quad \mathbf{H}_2 = \mathbf{y}A\frac{k}{k_s}\exp(ik_s z + \kappa_2 x), \quad (x < 0). \qquad (3)$$

Solution (2), (3) describes the double-evanescent wave with exponential amplitude decay on both sides from the interface, and the electric vector rotates in the propagation plane ($xz$) ("photonic wheel" [30,31]): clockwise at $x < 0$ and counter-clockwise at $x > 0$, seeing from the positive end of the $y$-axis (Fig. 1b shows the electric vector positions for a single moment of time but in different points of space). Here $\mathbf{x}$, $\mathbf{y}$ and $\mathbf{z}$ are the unit vectors of the coordinate axes, $A$ is the coordinate-independent normalization constant,

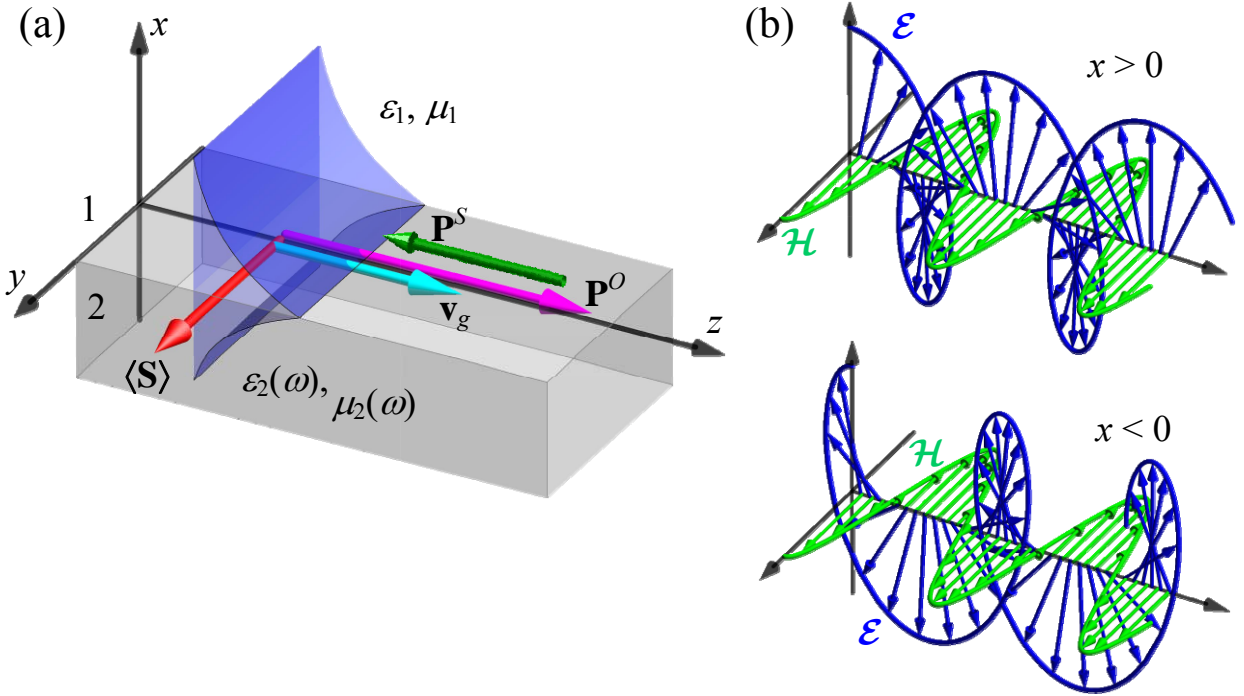

Fig. 1. (a) Geometrical configuration of a system supporting the SPP propagation; the cyan, magenta, red and green arrows show the group velocity (20) of the SPP field, orbital momentum (28), spin (25) and the volume part of the spin momentum (30), correspondingly (the singular surface part of the momentum (30) is not shown); (b) Instantaneous distributions of the electric and magnetic fields in the medium 1 ($x > 0$) and medium 2 ($x < 0$).

$$k_s^2 = \frac{\varepsilon_1 \varepsilon_2 (\varepsilon_1 \mu_2 - \varepsilon_2 \mu_1)}{\varepsilon_1^2 - \varepsilon_2^2} k^2, \quad \kappa_{1,2}^2 = \varepsilon_{1,2}^2 k^2 \frac{\varepsilon_2 \mu_2 - \varepsilon_1 \mu_1}{\varepsilon_1^2 - \varepsilon_2^2}, \quad \frac{\kappa_1}{\kappa_2} = -\frac{\varepsilon_1}{\varepsilon_2}. \tag{4}$$

As is seen, for existence of the TM mode, $\varepsilon_1$ and $\varepsilon_2$ must have opposite signs. For determinacy, we accept that $\varepsilon_1 > 0$, $\mu_1 > 0$, and then one of the two sets of conditions holds that enable the TM SPP propagation [4,5,10]: either

$$\varepsilon_2 < -\varepsilon_1, \quad \mu_2 > -\frac{\varepsilon_1 \mu_1}{|\varepsilon_2|} \tag{5a}$$

or

$$-\varepsilon_1 < \varepsilon_2 < 0, \quad \mu_2 < -\frac{\varepsilon_1 \mu_1}{|\varepsilon_2|}. \tag{5b}$$

### 3. Dynamical characteristics of the SPP field

3.1. Energy density is determined by the known Brillouin formula [29] which expresses the time-average energy stored in a medium with frequency-dependent permittivity and permeability:

$$w = \frac{g}{2} \left( \tilde{\varepsilon} |\mathbf{E}|^2 + \tilde{\mu} |\mathbf{H}|^2 \right) \tag{6}$$

where $g = (8\pi)^{-1}$. A special suitability of Eq. (6) is that it preserves the structure of the energy expression well known for simpler non-dispersive situations [28,29], $w_{ND} = (g/2)\left(\varepsilon |\mathbf{E}|^2 + \mu |\mathbf{H}|^2\right)$, and the medium dispersion is completely allowed for just by the replacement

$$\begin{Bmatrix} \varepsilon \\ \mu \end{Bmatrix} \to \begin{Bmatrix} \tilde{\varepsilon} \\ \tilde{\mu} \end{Bmatrix} = \frac{d}{d\omega}\left[\omega \begin{Bmatrix} \varepsilon(\omega) \\ \mu(\omega) \end{Bmatrix}\right]. \tag{7}$$

Importantly, $\tilde{\varepsilon}$ and $\tilde{\mu}$ are always positive,

$$\tilde{\varepsilon} > 0, \quad \tilde{\mu} > 0, \tag{8}$$

whereas $\varepsilon$ and $\mu$ can be negative (and this is typical for the conductive part of SPP-supporting structures, see Eqs. (5) and Fig. 1a).

In application to the TM field of Eqs. (2) and (3), Eq. (6) reduces to

$$w = \frac{g}{2}\left[\tilde{\varepsilon}\left(|E_x|^2 + |E_z|^2\right) + \tilde{\mu}|H_y|^2\right]. \tag{9}$$

Then, with account for the dispersion in the medium 2, this formula gives

$$w = \frac{g}{2}|A|^2 \begin{cases} \left[\dfrac{1}{\varepsilon_1}\left(1 + \dfrac{\kappa_1^2}{k_s^2}\right) + \mu_1 \dfrac{k^2}{k_s^2}\right] e^{-2\kappa_1 x} = \dfrac{2}{\varepsilon_1} e^{-2\kappa_1 x}, & x > 0; \\ \left[\dfrac{\tilde{\varepsilon}_2}{\varepsilon_2^2}\left(1 + \dfrac{\kappa_2^2}{k_s^2}\right) + \tilde{\mu}_2 \dfrac{k^2}{k_s^2}\right] e^{2\kappa_2 x} = \dfrac{2\varepsilon_2}{\varepsilon_1^2}\left[1 - \dfrac{1}{u}\dfrac{k}{k_s}\left(1 - \dfrac{\varepsilon_1^2}{\varepsilon_2^2}\right)\right] e^{2\kappa_2 x}, & x < 0 \end{cases} \tag{10}$$

where a dimensionless parameter $u$ is introduced by the relation

$$u = \frac{k}{k_s} \frac{\varepsilon_2^2 - \varepsilon_1^2}{\varepsilon_2^2 - \dfrac{\varepsilon_1^2 \varepsilon_2}{2}\left[\dfrac{\tilde{\varepsilon}_2}{\varepsilon_2^2}\left(1 + \dfrac{\kappa_2^2}{k_s^2}\right) + \tilde{\mu}_2 \dfrac{k^2}{k_s^2}\right]}. \tag{11}$$

At the interface $x = 0$, the energy distribution shows a discontinuity

$$\Delta w = w(+0) - w(-0) = g\frac{|A|^2}{\varepsilon_1}\left[1 - \frac{\varepsilon_2}{\varepsilon_1} + \frac{1}{u}\frac{k}{k_s}\frac{\varepsilon_2}{\varepsilon_1}\left(1 - \frac{\varepsilon_1^2}{\varepsilon_2^2}\right)\right]$$

$$= g\frac{|A|^2}{\varepsilon_1}\left(1 - \frac{\varepsilon_2}{\varepsilon_1}\right)\left(1 - \frac{1}{u}\frac{k}{k_s}\frac{\varepsilon_1 + \varepsilon_2}{\varepsilon_2}\right). \tag{12}$$

The total energy of the SPP can be determined by integration of the expression (10) over the whole space. Since the considered SPP field is homogeneous in both $z$- and $y$-directions, only the surface energy density is meaningful, which is determined by equation

$$\langle w \rangle = \int_{-\infty}^{\infty} w\, dx \tag{13}$$

(from now on, $\langle ... \rangle \equiv \int_{-\infty}^{\infty} ...\, dx$). Similarly to (13), the integral quantities can be introduced for other field characteristics: momentum, angular momentum, spin, etc.; we will call them "total" or "integral" keeping in mind that actually we deal with the surface densities across the interface plane. From Eq. (13) we immediately obtain

$$\langle w \rangle = \frac{g}{2}|A|^2 \left\{\frac{1}{\varepsilon_1 \kappa_1} + \frac{1}{2\kappa_2}\left[\frac{\tilde{\varepsilon}_2}{\varepsilon_2^2}\left(1 + \frac{\kappa_2^2}{k_s^2}\right) + \tilde{\mu}_2 \frac{k^2}{k_s^2}\right]\right\} \tag{14}$$

$$= g\frac{|A|^2}{2u}\frac{k}{k_s}\frac{1}{\varepsilon_1 \kappa_1}\left(1 - \frac{\varepsilon_1^2}{\varepsilon_2^2}\right). \tag{15}$$

3.2. Energy flow density is meaningfully described by the kinetic Abraham momentum [3,7,22,23], that is

$$\boldsymbol{\mathcal{I}} = c^2 \boldsymbol{\mathcal{P}}_A = gc\,\text{Re}\left[\mathbf{E}^* \times \mathbf{H}\right] = gc\,\text{Re}\left(-\mathbf{x}E_z^* H_y + \mathbf{z}E_x^* H_y\right) \tag{16}$$

(the second equality (16) is the result of application of the first one to the TM field of Section 1). Following [22,23], quantities associated with the kinetic momentum are highlighted by the calligraphic style, and the subscript "$A$" indicates the Abraham momentum definition; to the opposite, the Minkowski momentum and its relatives are not marked by special subscripts. The quantity (16) does not include the permittivities and/or permeabilities of the media, so it contains no explicit dependence on the medium dispersion. (This does not mean that the quantity (16) feels no dispersion but its influence is mediated by the field vectors $\mathbf{E}$ and $\mathbf{H}$). With employing Eqs. (2) – (4), Eq. (16) means

$$\boldsymbol{\mathcal{I}} = g|A|^2 \frac{\omega}{k_s} \mathbf{z} \begin{cases} \dfrac{1}{\varepsilon_1} e^{-2\kappa_1 x}, & x > 0; \\[2mm] \dfrac{1}{\varepsilon_2} e^{2\kappa_2 x}, & x < 0. \end{cases} \tag{17}$$

The energy flow discontinuity at $x = 0$ amounts to

$$\Delta \boldsymbol{\mathcal{I}} = \boldsymbol{\mathcal{I}}(+0) - \boldsymbol{\mathcal{I}}(-0) = \mathbf{z}g|A|^2 \frac{\omega}{k_s}\left(\frac{1}{\varepsilon_1} - \frac{1}{\varepsilon_2}\right). \tag{18}$$

The local velocity of the energy flow $\mathcal{I}_z/w$ is inhomogeneous (depends on $x$), as is seen from Eqs. (10) and (17). Moreover, the energy flows oppositely in media 1 and 2 [4,7]. The total quantity (17) $\langle \boldsymbol{\mathcal{I}} \rangle$, defined like (13), characterizes the SPP energy flow in the $z$-direction per unit $y$-width of the structure of Fig. 1a. It is easily obtained from Eqs. (17) and (4),

$$\langle \mathcal{I} \rangle = \mathbf{z} \langle \mathcal{I}_z \rangle = \mathbf{z} \frac{g}{2} |A|^2 \frac{\omega}{k_s} \frac{1}{\varepsilon_1 \kappa_1} \left(1 - \frac{\varepsilon_1^2}{\varepsilon_2^2}\right), \tag{19}$$

which enables us to calculate the group velocity of the whole SPP wave packet $\mathbf{v}_g = \mathbf{z} v_g$,

$$v_g = \frac{\langle \mathcal{I}_z \rangle}{\langle w \rangle} = cu \tag{20}$$

(see the cyan arrow in Fig. 1a; here, the group velocity coincides with the energy flow velocity due to the lossless character of the system but in presence of dissipation this equivalence may be violated [32]). According to Eq. (20), the parameter $u$ introduced in Eq. (11) is nothing but the SPP group velocity expressed in units of the vacuum light velocity. After some algebra involving Eqs. (4) and (7), one can show that the quantity (11), indeed, satisfies the general group velocity definition [28,29]

$$u = \frac{1}{c} \left(\frac{dk_s}{d\omega}\right)^{-1} = \frac{v_g}{c}. \tag{21}$$

Notably, under conditions (5a), the total energy flow $\langle \mathcal{I}_z \rangle$ and the group velocity $v_g$ are positive whereas under conditions (5b) they are negative ("backward" SPP [4]).

### 3.3. Spin density

The very suitable expression (6) of the electromagnetic energy in a dispersive medium is known for almost a century, and only recently, similar equations involving the dispersion-modified permittivity and permeability (7) were proposed for other dynamical characteristics [20–24]. In particular, it was found that the spin of the electromagnetic field in a dispersive medium is determined by the formula that follows from the Minkowski spin via the same substitution (7) as is made in the Brillouin formula (6) for the energy:

$$\mathbf{S} = \frac{g}{2\omega} \text{Im}\left(\tilde{\varepsilon} \mathbf{E}^* \times \mathbf{E} + \tilde{\mu} \mathbf{H}^* \times \mathbf{H}\right) = g \frac{\tilde{\varepsilon}}{2\omega} \mathbf{y} \, \text{Im}\left(E_z^* E_x - E_x^* E_z\right). \tag{22}$$

Being applied to Eqs. (2) and (3) with allowance for the non-dispersive character of the medium 1, Eq. (22) gives

$$\mathbf{S} = g \frac{|A|^2}{\omega k_s} \mathbf{y} \begin{cases} \frac{\kappa_1}{\varepsilon_1} e^{-2\kappa_1 x}, & x > 0; \\ -\frac{\tilde{\varepsilon}_2}{\varepsilon_2} \frac{\kappa_2}{\varepsilon_2} e^{2\kappa_2 x}, & x < 0. \end{cases} \tag{23}$$

Typically for the evanescent waves [7–10,15–19], the spin is directed orthogonally to the SPP propagation (red arrow in Fig. 1a). Like in the Abraham-based case [7,17], the spin directions in both media are opposite in agreement with the opposite senses of the vector $\mathbf{E}$ rotation in media 1 and 2 (see Fig. 1b). In general case, the SPP spin (23) is discontinuous at the interface $x = 0$:

$$\Delta \mathbf{S} = \mathbf{S}(+0) - \mathbf{S}(-0) = \mathbf{y} g \frac{|A|^2}{\omega} \frac{\kappa_1}{\varepsilon_1} \left(1 - \frac{\tilde{\varepsilon}_2}{\varepsilon_2}\right). \tag{24}$$

If the medium 2 were also non-dispersive, $\tilde{\varepsilon}_2 = \varepsilon_2$, $\Delta \mathbf{S} \equiv \Delta \mathbf{S}_{\text{ND}} = 0$: the "naïve" (dispersion-free) Minkowski spin is continuous [18].

The total spin of the SPP (in the sense of Eq. (13)) is determined via integration of the spin density (23) over the whole range of $x$, which, with account for Eqs. (19) and (15), gives

$$\langle \mathbf{S} \rangle = \mathbf{y} g \frac{|A|^2}{2\omega k_s \varepsilon_1} \left(1 - \frac{\tilde{\varepsilon}_2}{\varepsilon_2} \frac{\varepsilon_1}{\varepsilon_2}\right). \tag{25}$$

## 3.4. Orbital momentum density

Another dynamical characteristic that finds its natural expression with the dispersion-modified parameters (7) is the orbital (canonical) momentum [22,23],

$$\mathbf{P}^O = \frac{g}{2\omega} \text{Im}\left[ \tilde{\varepsilon} \mathbf{E}^* \cdot (\nabla) \mathbf{E} + \tilde{\mu} \mathbf{H}^* \cdot (\nabla) \mathbf{H} \right]. \quad (26)$$

For the TM field (2), (3), in view of relations

$$\text{Im}\left( \mathbf{E}^* \cdot \frac{\partial}{\partial x} \mathbf{E} \right) = \text{Im}\left( \mathbf{H}^* \cdot \frac{\partial}{\partial x} \mathbf{H} \right) = 0, \quad \frac{\partial}{\partial z} \begin{Bmatrix} \mathbf{E} \\ \mathbf{H} \end{Bmatrix} = ik_s \begin{Bmatrix} \mathbf{E} \\ \mathbf{H} \end{Bmatrix},$$

Eq. (26) leads to

$$\mathbf{P}^O = \mathbf{z} g \frac{k_s}{2\omega} \left[ \tilde{\varepsilon} \left( |E_x|^2 + |E_z|^2 \right) + \tilde{\mu} |H_y|^2 \right], \quad (27)$$

which means that the orbital momentum is always $z$-directed (see the magenta arrow in Fig. 1(a)) and distributed similarly to the energy distribution (9) – (12). Indeed, for the SPP field considered, both local and integral orbital momenta behave exactly as the local and integral energy and can be analyzed by Eqs. (10) – (12), (14) and (15):

$$\mathbf{P}^O = \mathbf{z} \frac{k_s}{\omega} w, \quad \langle \mathbf{P}^O \rangle = \mathbf{z} \frac{k_s}{\omega} \langle w \rangle = \frac{\langle \mathcal{I} \rangle}{c^2 u} \frac{k_s}{k} \quad (28a)$$

or

$$\mathbf{P}^O = \mathbf{z} g \frac{|A|^2}{\omega} k_s \begin{cases} \dfrac{1}{\varepsilon_1} e^{-2\kappa_1 x}, & x > 0 ; \\ \dfrac{\varepsilon_2}{\varepsilon_1^2} \left[ 1 - \dfrac{1}{u} \dfrac{k}{k_s} \left( 1 - \dfrac{\varepsilon_1^2}{\varepsilon_2^2} \right) \right] e^{2\kappa_2 x}, & x < 0. \end{cases} \quad (28b)$$

As $k_s/k > 1$, Eqs. (28) demonstrate the known fact that the orbital momentum in an evanescent wave exceeds that of the plane wave with the same energy and frequency [7,8,33,34].

## 3.5. Spin momentum density

The spin momentum (Belinfante momentum) [7,8,25,26] of the electromagnetic field in a dispersive medium is directly related with the dispersion-modified spin (22) [24]:

$$\mathbf{P}^S = \frac{1}{2} \nabla \times \mathbf{S} = \frac{1}{2} \left( -\mathbf{x} \frac{\partial S_y}{\partial z} + \mathbf{z} \frac{\partial S_y}{\partial x} \right) \quad (29)$$

(the second equality (29) follows from the first one with allowance for the SPP geometry). Taking Eq. (23) into account we obtain

$$\mathbf{P}^S = \mathbf{z} g \frac{|A|^2}{\omega k_s} \begin{cases} -\dfrac{\kappa_1^2}{\varepsilon_1} e^{-2\kappa_1 x}, & x > 0 \\ -\dfrac{\tilde{\varepsilon}_2}{\varepsilon_2} \dfrac{\kappa_2^2}{\varepsilon_2} e^{2\kappa_2 x}, & x < 0 \end{cases} + \mathbf{P}^{\text{surf}}, \quad \mathbf{P}^{\text{surf}} = \mathbf{z} g \frac{|A|^2}{2\omega k_s} \frac{\kappa_2}{\varepsilon_2} \left( \frac{\tilde{\varepsilon}_2}{\varepsilon_2} - 1 \right) \delta(x). \quad (30)$$

The spin momentum consists of the volume part (in curly brackets) and the surface contribution $\mathbf{P}^{\text{surf}}$ described by the delta-function, which formally appears because of the spin discontinuity (24). This singular term owes to the dispersion and vanishes if $\tilde{\varepsilon}_2$ is replaced by $\varepsilon_2$. In view of Eqs. (5) and (8), the volume contribution of Eq. (30) is directed oppositely to the SPP propagation (green arrow in Fig. 1a) while the singular surface term describes the positively directed momentum. Importantly, the volume and surface contributions of the spin momentum exactly compensate each other so that the total spin momentum integrated over the whole space vanishes, as is required by the general theory [26,27]:

$$\langle \mathbf{P}^S \rangle = \int_{-\infty}^{\infty} \mathbf{P}^S dx = 0. \tag{31}$$

Equations (28) and (30) enable us to find the dispersion-modified kinetic Minkowski-type momentum as a sum of the spin and orbital momenta:

$$\mathcal{P} = \mathbf{P}^S + \mathbf{P}^O = \mathbf{z} g \frac{|A|^2}{2c} \frac{k}{k_s} \left[ \left\{ \begin{array}{ll} 2\mu_1 e^{-2\kappa_1 x}, & x > 0 \\ \frac{\tilde{\varepsilon}_2 \mu_2 + \varepsilon_2 \tilde{\mu}_2}{\varepsilon_2} e^{2\kappa_2 x}, & x < 0 \end{array} \right\} + \frac{\kappa_2}{\varepsilon_2 k^2}\left(\frac{\tilde{\varepsilon}_2}{\varepsilon_2} - 1\right)\delta(x) \right]. \tag{32}$$

### 3.6. Orbital angular momentum of the SPP

The orbital angular momentum (OAM) is, generally, an extrinsic dynamical characteristic of an electromagnetic field that depends on the reference point $\mathbf{r}_0$ [35]. Its density is defined by equation [27,35]

$$\mathbf{L} = (\mathbf{r} - \mathbf{r}_0) \times \mathbf{P}^O = -\mathbf{y}(x - x_0)P_z^O \tag{33}$$

(here, as usual in sections 3.2 – 3.5, the first equality (33) expresses the general definition and the second one is its realization for the considered TM field: as the orbital momentum (28) is always $z$-directed, and the SPP field is homogeneous along the $y$-direction, only the $x$-component of the momentum arm $\mathbf{r} - \mathbf{r}_0$ is influential). Eq. (33) testifies that the OAM is, generally, not an independent characteristic of the SPP field but is governed by the orbital momentum (28) or the energy (10) distribution. However, its mechanical action can be important and separately observed in some cases [7,22,36,37].

According to Eq. (33), the total (in the sense of Eq. (13)) OAM of the SPP is determined by equation (see Eq. (28a))

$$\langle \mathbf{L} \rangle = \int_{-\infty}^{\infty}(\mathbf{r} - \mathbf{r}_0) \times \mathbf{P}^O dx = -\mathbf{y}\frac{k_s}{\omega}\int_{-\infty}^{\infty}(x - x_0)w\,dx = -\mathbf{y}\frac{k_s}{\omega}(\langle x \rangle - x_0)\langle w \rangle \tag{34}$$

where

$$\langle x \rangle = \frac{1}{\langle w \rangle}\int_{-\infty}^{\infty} x w\, dx \tag{35}$$

is the "energy center" of the SPP known as a geometric characteristic of the energy distribution (9), (10) along the axis $x$ normal to the interface [19,22]. With using Eqs. (10) and (11), one finds

$$\langle x \rangle = \frac{1}{2\kappa_1}\left(\frac{\varepsilon_1}{\varepsilon_2} + u\frac{k_s}{k}\frac{\varepsilon_2}{\varepsilon_2 + \varepsilon_1}\right). \tag{36}$$

The energy center singles out a "privileged" reference point associated with the SPP field 'in itself' rather than with its "occasional" position within an external coordinate frame. Therefore, the OAM with respect to the energy center $x_0 = \langle x \rangle$ is usually referred to as 'intrinsic'; according to Eq. (34), for the SPP field this intrinsic OAM vanishes, $\langle \mathbf{L}^{int} \rangle = 0$. Another meaningful choice of the reference point is to measure the OAM with respect to the interface $x_0 = 0$; this OAM is sometimes called 'extrinsic' $\langle \mathbf{L}^{ext} \rangle$ in the narrow sense [35]. Following to Eqs. (34), (36) and (15), it can be determined as

$$\langle \mathbf{L}^{ext} \rangle = -\mathbf{y}\frac{k_s}{\omega}\langle x \rangle\langle w \rangle = -\mathbf{y} g \frac{|A|^2}{4c}\frac{1}{\varepsilon_1 \kappa_1^2}\left(1 - \frac{\varepsilon_1}{\varepsilon_2}\right)\left[\frac{\varepsilon_1}{u\varepsilon_2}\left(1 + \frac{\varepsilon_1}{\varepsilon_2}\right) + \frac{k_s}{k}\right]. \tag{37}$$

3.7. Imaginary Poynting momentum

For completeness, we additionally present the expressions for the imaginary Poynting momentum (reactive momentum [28]) that also plays an important role in the light-matter interaction [38], in particular, it is responsible for the noticeable ponderomotive action on material particles [8,39,40]. Like the energy flow density (16), it is an essentially kinetic property (the physical meaning of its spin-orbital decomposition is doubtful) and shows no explicit dependence on dispersion (more exactly, its known physical manifestations are not related to the dispersion). The following equations illustrate both the Minkowski and Abraham versions of this quantity for the SPP field:

$$\mathbf{\Pi} = \varepsilon\mu\mathbf{\Pi}_A = \frac{g}{c}\text{Im}\left[\varepsilon\mu\mathbf{E}^* \times \mathbf{H}\right] = g|A|^2 \frac{k}{ck_s^2}\mathbf{x}\begin{cases}-\mu_1\kappa_1 e^{-2\kappa_1 x}, & x > 0; \\ \mu_2\kappa_2 e^{2\kappa_2 x}, & x < 0.\end{cases}$$

Note that due to the last Eq. (4) the Abraham-based reactive momentum is continuous at the boundary $x = 0$.

### 4. SPP-induced magnetization

As was shown by the microscopic analysis of a simpler case [22–24], the analog of the surface (singular) term of Eq. (30) was caused by the directional motion of free electrons in the metallic medium 2 connected to its magnetization due to the inverse Faraday effect [29,41]. This conclusion was made based on the specific model of the medium 2. Now we try to consider the similar problem with minimum special assumptions relating the medium 2 properties, mainly grounding on its parameters $\varepsilon_2(\omega)$, $\mu_2(\omega)$ and their dispersion-modified derivatives (7).

The results of Section 2 testify that, as a rule, the dynamical characteristics behave regularly at the metal-dielectric interface $x = 0$: Eqs. (10), (12), (17), (18), (23), (24) and (28) show possible discontinuities but no singular terms which are hardly interpretable. The only exclusion is the spin momentum (30) where the delta-term cannot be ascribed to any feature of the electromagnetic field 'per se'. Keeping in mind the microscopic arguments of [3,22,23], it is natural to associate the delta-function term of Eq. (30) with the surface motion of the free charge carriers. If the charge carriers are electrons with charge $e < 0$ and mass $m$, the singular surface momentum $\mathbf{P}^{\text{surf}}$ of Eq. (30) indicates the existence of the corresponding surface electric current with the density

$$\mathbf{j}^{\text{surf}} = \frac{e}{m}\mathbf{P}^{\text{surf}} = \mathbf{z}g\frac{|A|^2}{\omega k_s}\frac{e}{2m}\frac{\kappa_2}{\varepsilon_2}\left(\frac{\tilde{\varepsilon}_2}{\varepsilon_2} - 1\right)\delta(x). \tag{38}$$

This current, strictly localized at the surface of the conductive medium, can be considered as an analog of the Ampere current explaining the magnetization of permanent magnets [42]. In particular, due to the Maxwell's boundary condition [28,29,42], the surface current (38) means that the magnetic field discontinuity occurs at the interface

$$B_y(+0) - B_y(-0) = \frac{4\pi}{c}\int_{-\infty}^{\infty} j_z^{\text{surf}} dx. \tag{39}$$

This is a time-independent (DC) magnetic field that is not described by Eqs. (1) – (3), and its emergence can be treated as a sort of non-linear interaction between the SPP light wave and the conductive medium 2. Since there are no free charges in the dielectric medium 1, the result (39) implies that a DC magnetic field is expected to exist only in the medium 2. Actually, it shows that, in the course of the SPP propagation, the DC magnetic field and, consequently, the permanent magnetization is induced in the conductive part of the SPP-supporting structure (Fig. 1a). Eq. (39) enables to specify its near-boundary value

$$\mathbf{B}(0) = -\mathbf{y}\frac{4\pi}{c}\int_{-\infty}^{\infty} j_z^{\text{surf}} dx = \mathbf{y}g\frac{4\pi|A|^2}{\omega k_s}\frac{-e}{2mc}\frac{\kappa_2}{\varepsilon_2}\left(\frac{\tilde{\varepsilon}_2}{\varepsilon_2} - 1\right). \tag{40}$$

In principle, relationship (40) is sufficient for calculation of the DC magnetic field across the whole half-space $x < 0$ because one can naturally assume that $\mathbf{B}(x)$ decays with the off-interface distance similarly to all other characteristics,

$$\mathbf{B}(x) = \mathbf{B}(0)\exp(2\kappa_2 x). \tag{41}$$

However, now we consider a bit different reasoning which additionally elucidates the physical meaning of some quantities involved in the problem.

Evidently, the surface current (38) should be a part of a closed circuit that can be completed by the equivalent oppositely directed current density in the medium 2 volume, $\mathbf{j}_{\text{magn}}$, related to the volume magnetization $\mathbf{M} = \mathbf{y} M_y$:

$$\mathbf{j}_{\text{magn}} = c\nabla \times \mathbf{M} = \mathbf{z} c \frac{\partial M_y}{\partial x}. \tag{42}$$

Both $\mathbf{j}^{\text{surf}}$ and $\mathbf{j}_{\text{magn}}$ are bound currents [42,43] that exist without external electric voltage, do not contribute to the charge transport and cannot be measured by an ammeter [23,42]. The integral value of the current (42) must compensate the integral contribution of the surface term (38):

$$\langle \mathbf{j}_{\text{magn}} \rangle = \int_{-\infty}^{0} \mathbf{j}_{\text{magn}} dx = \frac{e}{m} \int_{-\infty}^{0} \mathbf{P}_{\text{magn}} dx = \mathbf{z} g \frac{|A|^2}{\omega k_s} \frac{-e}{2m} \frac{\kappa_2}{\varepsilon_2} \left( \frac{\tilde{\varepsilon}_2}{\varepsilon_2} - 1 \right)$$

where $\mathbf{P}_{\text{magn}}$ is the mechanical momentum associated with the volume magnetization current $\mathbf{j}_{\text{magn}}$ (its analog in [22,23] was called "magnetization momentum"). Since, like all other characteristics, it is distributed proportionally to $\exp(2\kappa_2 x)$, this results in the volume current expression

$$\mathbf{j}_{\text{magn}} = \frac{e}{m} \mathbf{P}_{\text{magn}} = \mathbf{z} g \frac{|A|^2}{\omega k_s} \frac{-e}{m} \frac{\kappa_2^2}{\varepsilon_2} \left( \frac{\tilde{\varepsilon}_2}{\varepsilon_2} - 1 \right) e^{2\kappa_2 x}.$$

Hence, Eq. (42) with the natural condition $\mathbf{M}(x)|_{x \to -\infty} = 0$ gives an expression for the medium 2 magnetization:

$$\mathbf{M} = g \frac{|A|^2}{\omega k_s} \mathbf{y} \frac{-e}{2mc} \frac{\kappa_2}{\varepsilon_2} \left( \frac{\tilde{\varepsilon}_2}{\varepsilon_2} - 1 \right) e^{2\kappa_2 x}, \tag{43}$$

which agrees with Eqs. (40) and (41) in view of the relation $\mathbf{B} = 4\pi \mathbf{M}$ that holds in the unlimited medium 2 without external magnetic field [29,42] (note, however, that this gives no direct indications concerning the observable magnetic field near a real metal strip of a finite $y$-width that is modeled by the unlimited medium 2).

According to Eq. (43), the integral SPP-induced magnetization (the magnetic moment of the conductive-medium layer per unit $z$-length and unit $y$-width) equals to

$$\langle \mathbf{M} \rangle = \int_{-\infty}^{0} \mathbf{M} \, dx = \mathbf{y} g \frac{|A|^2}{2\omega k_s} \frac{-e}{2mc} \frac{\tilde{\varepsilon}_2 - \varepsilon_2}{\varepsilon_2^2}. \tag{44}$$

Noteworthy, the magnetization effect is of essentially dispersive nature: it vanishes when $\tilde{\varepsilon}_2 = \varepsilon_2$. This is expectable because the magnetization occurs due to the same properties of the medium 2 (the electron gas susceptibility to the external field) that determine its permeability. This correspondence can be continued. In particular, the magnetization, being coupled with mechanical motion of the electrons, can be associated with the mechanical momentum via the standard gyromagnetic ratio

$$\mathbf{S}_{\text{magn}} = \frac{2mc}{e} \mathbf{M} = g \frac{|A|^2}{\omega k_s} \mathbf{y} \frac{\kappa_2}{\varepsilon_2} \left( 1 - \frac{\tilde{\varepsilon}_2}{\varepsilon_2} \right) e^{2\kappa_2 x}.$$

This mechanical angular momentum can be interpreted as the material contribution to the SPP spin (23) (cf. Eq. (4.25) of [22]); accordingly, the magnetization momentum $\mathbf{P}_{magn} = (1/2)\nabla \times \mathbf{S}_{magn}$ can be treated as the material contribution to the total spin momentum $\mathbf{P}^S$ (30). Interestingly, after exclusion of the material parts $\mathbf{S}_{magn}$ and $\mathbf{P}_{magn}$, the remaining "field" parts of the spin and spin momentum in the medium 2,

$$\mathbf{S} - \mathbf{S}_{magn} = -\mathbf{y}g\frac{|A|^2}{\omega k_s}\frac{\kappa_2}{\varepsilon_2}e^{2\kappa_2 x}, \quad \mathbf{P}^S - \mathbf{P}_{magn} = -\mathbf{z}g\frac{|A|^2}{\omega k_s}\frac{\kappa_2^2}{\varepsilon_2}e^{2\kappa_2 x}, \; x < 0,$$

acquire a remarkable similarity to the "pure field" contributions in the $x > 0$ half-space (cf. Eqs. (23), (30)) and coincide with the "naïve" Minkowski-based expressions for $\mathbf{S}$ and $\mathbf{P}^S$.

## 5. Discussion and illustrations

The main results of the above sections are explicit analytical formulas characterizing spatial distributions of the electromagnetic energy (Eqs. (10), (15), (36)), energy flow (Eqs. (17), (19)) and its group velocity (11), orbital (28) and spin (30) parts of the electromagnetic momentum, spin (23), (25) and orbital (34), (37) angular momenta in the TM SPP fields. Additionally, the SPP-induced magnetization of the conductive medium 2 is phenomenologically described by Eqs. (43), (44).

Since the main attention is paid to the spatial inhomogeneity of the field characteristics, their absolute values are not well defined by the presented equations and are expressed through the arbitrarily chosen constant $A$ related with the amplitude of the electric field normal to the interface (see Eqs. (2) and (3)). However, this choice is by no means unique and can be inconvenient if, for example, the frequency-dependent behavior of the SPP dynamical characteristics is of interest. In many such situations, the SPP energy flow $\langle \mathcal{I}_z \rangle$ (19) or (directly associated with it) energy (15), (20) are the quantities that can be efficiently controlled and measured, rather than the transverse electric field amplitude chosen in Eqs. (2) and (3). Then the main analytical results of this paper can be customized by the replacements

$$g|A|^2 = \frac{2k_s}{\omega}\langle \mathcal{I}_z \rangle \varepsilon_1 \kappa_1 \frac{\varepsilon_2^2}{\varepsilon_2^2 - \varepsilon_1^2} = \frac{2k_s}{k}u\langle w \rangle \varepsilon_1 \kappa_1 \frac{\varepsilon_2^2}{\varepsilon_2^2 - \varepsilon_1^2}. \tag{45}$$

In the forms resulting from the first equality (45), the explicit functions $\varepsilon_2(\omega)$, $\mu_2(\omega)$, $\kappa_{1,2}(\omega)$ and $k_s(\omega)$, known from the phenomenological data or from relations (4), immediately express the frequency-dependent SPP characteristics for the given energy flow (SPP power). Alternatively, the dynamical quantities per unit SPP energy ('per photon') can be easily found via the second relation (45) or its equivalent (15).

To illustrate the results we employ two popular examples of the SPP-supporting structures where the medium 1 is vacuum ($\varepsilon_1 = \mu_1 = 1$) and the medium 2 is formed by noble metals Ag or Au. In both cases $\mu_2(\omega) = 1 =$ const but the dielectric function $\varepsilon_2(\omega)$ shows a complicated dependence on frequency. Fig. 2a presents the data of [44] for the frequency region with minimal dissipation (following to [44], in all graphs of Fig. 2 the frequency is expressed via the corresponding values of the quantum of energy $\hbar\omega$; the "genuine" frequency in $s^{-1}$ can be found from the "energy" data of Fig. 2 via multiplication by $\hbar^{-1} = 1.52 \cdot 10^{15} \, \mathrm{eV}^{-1}\mathrm{s}^{-1}$). It is seen from Fig. 2a that the imaginary parts of the permittivity (dashed curves) are not very small; however, within the frame of the present work, they are neglected and the SPP field is calculated under assumption that for both metals permittivities are real. The calculations are restricted to the frequency range with the lower limit determined by increase of absorption (see [44]) and the upper limit dictated by the condition (5a) according to which the propagating SPP can only exist while $\mathrm{Re}\,\varepsilon_2 < -1$. This means that $\omega$ does not exceed the cut-off frequency $\omega_c$: for Ag $\hbar\omega_c = 3.595$ eV, for Au $\hbar\omega_c = 2.820$ eV. For the sake of comparison, there are also presented data calculated for the SPM [5,22–24] with $\varepsilon_2(\omega) = 1 - \omega_p^2/\omega^2$ (thin blue curves) where $\omega_p$ was chosen such that the cutoff frequency of the SPM, $\omega_p/\sqrt{2}$, coincide with the calculated cut-off frequency for silver ($\hbar\omega_p = 5.084\,\text{eV}$) and for gold ($\hbar\omega_p = 3.988\,\text{eV}$) – these values follow from the data of [44] reproduced in Fig. 2a as points where the thick curves cross the horizontal level $-\text{Re}\varepsilon_2(\omega) = 1$. The curve $\text{Im}\varepsilon_2(\omega)$ for the SPM is not shown in Fig. 2a because in this case $\text{Im}\varepsilon_2(\omega) = 0$.

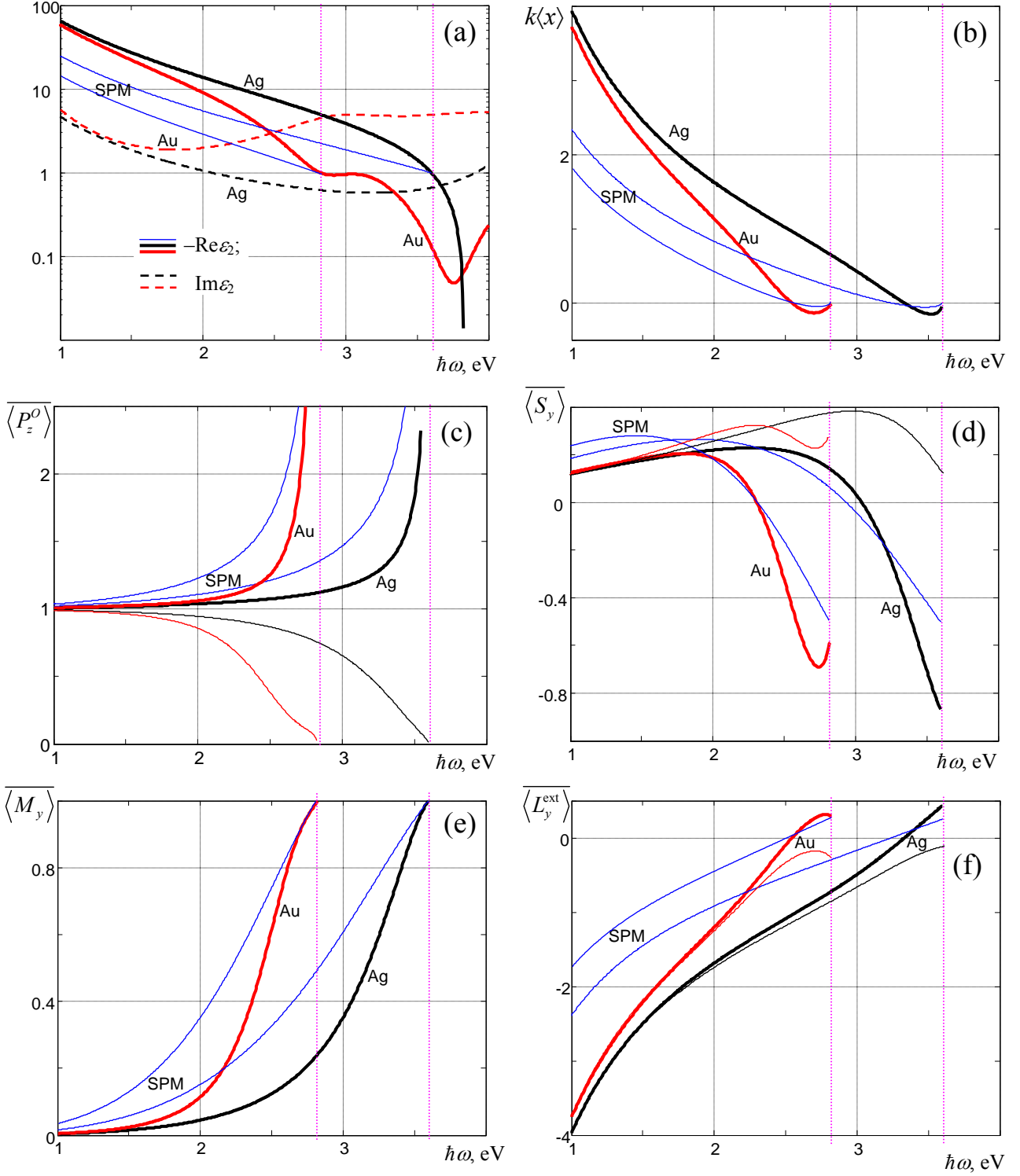

Fig. 2. (a) Optical constants of the medium 2 and (b) – (f) dynamical characteristics of the SPP field formed when the medium 1 is vacuum while the medium 2 is silver (black curves, Ag) and gold (red curves, Au): (b) normalized energy center position (36); (c) orbital momentum (28a), (46); (d) spin (25), (46); (e) magnetization (44), (47); (f) OAM (37), (46). Thin black and red curves in (c), (d) and (f) illustrate the case when $\tilde{\varepsilon} \to \varepsilon$ in Eqs. (25) and (26), (27), (33); for comparison, the results obtained if the medium 2 is described by the SPM [5, 22–24] are also presented by thin blue curves marked "SPM". The frequency is scaled in units of the corresponding quantum of energy $\hbar\omega$; vertical dotted lines mark the cutoff frequencies at which relation (5a) violates.

Fig. 2b demonstrates the frequency-dependent behavior of the energy center (35) calculated by Eq. (36) and expressed in units of $k^{-1} = c/\omega$. The orbital momentum (28a), spin (25) and extrinsic OAM (37) (Figs. 2c, d and f) are presented in dimensionless normalized forms

$$\overline{\langle P_z^O \rangle} = \frac{c\langle P_z^O \rangle}{\langle w \rangle}, \quad \overline{\langle S_y \rangle} = \frac{\omega\langle S_y \rangle}{\langle w \rangle}, \quad \overline{\langle L_y \rangle} = \frac{\omega\langle L_y \rangle}{\langle w \rangle} \qquad (46)$$

whereas the magnetization (44) 'per plasmon' (Fig. 2e) is expressed in units of the Bohr magneton [45],

$$\overline{\langle M_y \rangle} = \frac{\langle M_y \rangle \hbar\omega}{\langle w \rangle (-\hbar e/2mc)}. \qquad (47)$$

For comparison, images (c), (d) and (f) also illustrate the dependences expected without the dispersion corrections of [22–24]. Thin red (Au) and black (Ag) curves are calculated via Eq. (25) for (d) and directly by Eqs. (26), (33) for (c) and (f), assuming $\tilde{\varepsilon}_2 = \varepsilon_2$ instead of Eq. (7), describe the behavior of the orbital linear momentum and of the spin and orbital angular momenta determined by using the traditional ("naïve") Minkowski momentum definition. It is seen that the dispersion-modified quantities (thick curves) remarkably differ from the "naïve" Minkowski-based prototypes. The role of dispersion corrections can be negligible in the low-frequency region (where, due to relation $|\varepsilon_2/\varepsilon_1| \gg 1$, the SPP field is mainly concentrated in the non-dispersive medium 1, cf. Fig. 2b) but grows near the cut-offs. Their influence becomes especially impressive for the SPP canonical momentum and spin (Figs. 2c, d); the discrepancies between the dispersion-modified and "naïve" OAMs in Fig. 2f are relatively small because the energy center is close to the interface $x = 0$ in the near-cutoff region (cf. Fig. 2b and Eq. (37)). The positive sign of the spin in Fig. 2d means that the direction of propagation **z**, outer normal to the metal surface **x** and the spin vector form a right-handed system; such an orientation of the spin (and the opposite one of the OAM) agrees with multiple earlier calculations (e.g., [7,10,15–19]) performed for the interfaces between "normal" metals and dielectrics. In this view, it is remarkable that the dispersion-modified angular momenta in Figs. 2(d) and 2(f) change the sign in the considered frequency range while the "naïve" ones do not.

The same assumption $\tilde{\varepsilon}_2 = \varepsilon_2$ leads to zero magnetization (44). As was already said in Section 4, without the dispersion corrections $\langle M_y \rangle$ vanishes, and that is why the expected thin red and black curves are not shown in Fig. 2e.

All curves of Figs. 2b–f demonstrate remarkable deviations from the simplified results based on the SPM (blue curves). However, the differences are mainly quantitative, and, in general, main features characterising the frequency dependence of the SPP dynamical characteristics depicted in Figs. 2b–f are reflected by the SPM model. As a general rule, one can notice that if the medium 2 is formed by a real metal, all the characteristics presented in Figs. 2b–f show weaker variations far from the cut-off frequency but change more rapidly in the near-cutoff region (as compared to the SPM-based data, for which the variations are more smooth and uniform). The only qualitative difference is a slight decrease of the absolute values of the spin and OAM for the SPP supported by the golden film (crooks of the red curves in panels (d) and (f) near the cut-off), which can be ascribed to the rapid decrease of |Re$\varepsilon_2$ ($\omega$)| in the corresponding frequency region (see the red curve in panel (a)).

It can be expected, however, that such relatively weak deviations from the conclusions based on the SPM take place only for usual metallic media 2; in more complex structures including metamaterials with arbitrary $\varepsilon_2$ ($\omega$), $\mu_2$ ($\omega$) showing intricate frequency dependences [14,46,47], the behavior of dynamical characteristics will be much more interesting and rich of details. The mathematical expressions developed in Section 3 supply adequate analytical means for such cases.

### 6. Concluding remarks

The main content of this paper represents a collection of formulas expressing the dynamical characteristics of the SPP fields obtained with application of recent results for the momentum and angular momentum in dispersive media [22–24]. These formulas constitute a coherent system of relations consistently describing the spatial distributions of the electromagnetic energy, momentum and spin of the SPP fields, thus providing a set of analytical instruments which can be useful in various SPP applications.

Importantly, the dynamical characteristics' representations obtained demonstrate a complete consistency and absence of hardly interpretable singularities; the canonical (spin-orbital) decomposition of the field momentum in a highly inhomogeneous SPP-supporting system is performed without additional terms caused by the medium inhomogeneity that are typical for earlier attempts based on the Abraham approach [7,17,48]. The possibility of the 'neat' spin-orbital decomposition under spatially inhomogeneous conditions is, in fact, a feature of the Minkowski momentum [22–24,49], and it is inherited, in part, by the dispersion-modified quantities. Moreover, the only exclusion – the singular surface spin momentum (30) – appears, specifically, due to explicit account for the dispersion and contributes to another general result of the paper: the phenomenological derivation of the SPP-induced magnetization in Section 4.

The magnetization accompanying the SPP propagation, as a sort of inverse Faraday effect, was first predicted based on the Drude model of the metal medium 2 [22,23]. In this work, the only assumption is that the charge carriers in the medium 2 are electrons with given mass and charge, and the magnetization appears as a natural consequence of the permittivity dispersion in the medium 2. It is quite expectable that the similar reasoning can be applied to cases where the charge carriers are quasiparticles with different effective charges and masses, which is typical to metamaterials [14,46,47]. However, a reasonable association of the surface momentum in Eq. (30) with a certain sort of electric current can only be made on the base of a specific model relating the nature of the charge carries in the medium 2. E.g., a mechanical momentum of the common motion of oppositely charged and equally massive quasiparticles means no electric current (instead of the surface current (38)) and, correspondingly, no magnetization.

All the above results are obtained for the TM fields but they can be extended to the TE modes that are also possible in the SPP-supporting structure depicted in Fig. 1a [4,10,15]. Formally, the TE modes are electromagnetically dual [28] to the TM solutions expressed by Eqs. (2) – (5), and formulas for the TE field can be obtained from their TM counterparts via substitutions

$$\mathbf{H} \to \mathbf{E}, \quad \mathbf{E} \to -\mathbf{H}, \quad \varepsilon \rightleftarrows \mu. \qquad (48)$$

In particular, for the geometry of Fig. 1a, the electric field of the TE mode is represented by the only $y$-component with continuous at $x = 0$ amplitude $A(k/k_s)$ while the magnetic field in both media rotates, and its $x$-components near the interface are $A/\mu_1$ and $A/\mu_2$ (cf. Eqs. (2) and (3)). Note that the TE analogs of Eqs. (5) determining the SPP existence are incompatible with the 'original' Eqs (5): the TE and TM modes cannot be supported simultaneously (except the degenerate case $\varepsilon_2 = \mu_2 = -\varepsilon_1 = -\mu_1$) [4,10].

In formulas of Section 3 that do not contain $\mathbf{E}$ and $\mathbf{H}$ explicitly, the mutual replacement $\varepsilon \rightleftarrows \mu$ is sufficient for transition to the TE case; however, Section 4 treating the SPP-induced magnetization essentially employs the physical interpretation applicable only to TM fields. Formally, in the TE modes, a singular surface momentum similar to Eq. (30) also appears,

$$\left(\mathbf{P}^{\text{surf}}\right)_{TE} = \mathbf{z} g \frac{|A|^2}{2\omega k_s} \frac{\kappa_2}{\mu_2}\left(\frac{\tilde{\mu}_2}{\mu_2} - 1\right) \delta(x) \tag{49}$$

(where $k_s$ and $\kappa_2$ are determined by modified Eqs. (4) with interchanged permittivities and permeabilities) but its interpretation is not as direct as that given by Eq. (38). We cannot associate expression (49) with the surface electric current and/or magnetization without additional knowledge on the medium 2 structure which, in this case, is a complex metamaterial with the dispersive permeability $\mu_2(\omega) < 0$.

The unified systematic approach to the energy and momentum characteristics of the SPP fields that is described in this work and originates from [22–24] is not free from deficiencies. For example, a remarkable flexibility and efficient control of the SPP properties can be realized in configurations with anisotropic and/or spatially inhomogeneous dielectric medium (see, e.g., [16] and references therein). In such situations, the basic equations for the SPP field are more complicated than Eqs. (2) – (4) but once the electric and magnetic fields are found analytically or numerically, the dynamical characteristics of the SPP field can be evaluated following to the same scheme. The most important disadvantage of the presented approach is that it does not include the energy dissipation (all $\varepsilon_{1,2}$ and $\mu_{1,2}$ are supposed to be real). As is seen from Fig. 2a, in typical practical situations this is a rather rough approximation, and any further development of the theory must include its generalization to lossy media. However, this is an inherent drawback of any macroscopic approach: there exist fundamental limitations for a phenomenological description of the field momentum and angular momentum in media with dissipation [29]. Most probably, the dynamical characteristics of the SPP field in dissipative media can be described on the microscopic ground with employing specific models of the medium structure, known properties of the charge carriers, etc. Nevertheless, development of a consistent phenomenological description for the electromagnetic momentum (and, more generally, electromagnetic stress tensor) in presence of dissipation is still a task for future research.

### Acknowledgments

The authors thank Konstantin Bliokh from RIKEN, Wako-shi, Japan, for fruitful discussions and valuable assistance. This work was supported by the Ministry of Education and Science of Ukraine, project No 582/18 (State Registration Number 0118U000198).

### References

1. M. I. Stockman et al, Roadmap on plasmonics, J. Opt. 20 (2018) 043001.
2. L. Novotny, B. Hecht, Principles of Nano-Optics, Cambridge University Press, Cambridge, 2012.
3. J. Nkoma, R. Loudon, D. R. Tilley, Elementary properties of surface polaritons, J. Phys. C: Solid State Phys. 7 (1974) 3547–3559.
4. I. V. Shadrivov, A. A. Sukhorukov, Y. S. Kivshar, A. A. Zharov, A. D. Boardman, P. Egan, Nonlinear surface waves in left-handed materials, Phys. Rev. E 69 (2004) 016617.
5. A. V. Zayats, I. I. Smolyaninov, A. A. Maradudin, Nano-optics of surface plasmon polaritons, Phys. Rep. 408 (2005) 131–314.
6. J. P. B. Mueller, F. Capasso, Asymmetric surface plasmon polariton emission by a dipole emitter near a metal surface, Phys. Rev. B 88 (2013) 121410.
7. K. Y. Bliokh, F. Nori, Transverse spin of a surface polariton, Phys. Rev. A 85 (2012) 061801.
8. K. Bliokh, A. Bekshaev, F. Nori, Extraordinary momentum and spin in evanescent waves, Nat. Commun. 5 (2014) 3300.

9. J. Lin, J. P. B. Mueller, Q. Wang, G. Yuan, N. Antoniou, X.-C. Yuan, F. Capasso, Polarization-controlled tunable directional coupling of surface plasmon polaritons, Science 340 (2013) 331–334.
10. K. Y. Bliokh, D. Smirnova, F. Nori, Quantum spin Hall effect of light, Science 348 (2015) 1448–1451.
11. R. E. Camley, Nonreciprocal surface waves, Surf. Sci. Rep. 7 (1987) 103–187.
12. K. Y. Bliokh, F. J. Rodríguez-Fortuño, A. Y. Bekshaev, Y. S. Kivshar, F. Nori, Electric-current-induced unidirectional propagation of surface plasmon-polaritons, Opt. Lett. 43 (2018) 963–966.
13. D. R. Smith, W. Padilla, D. C. Vier, S. C. Nemat-Nasser, S. Schultz, Composite medium with simultaneously negative permeability and permittivity, Phys. Rev. Lett. 84 (2000) 4184–7418.
14. W. J. Padilla, D. N. Basov, D. R. Smith, Negative refractive index metamaterials, Materials Today 9 (2006) 28–35.
15. R. Ruppin, Surface polaritons of a left-handed medium, Phys. Lett. A 277 (2000) 61–64.
16. X. Xiao, M. Faryad, A. Lakhtakia, Multiple trains of same-color surface-plasmon-polaritons guided by the planar interface of a metal and a sculptured nematic thin film. Part VI: Spin and orbital angular momentums, J. Nanophoton. 7 (2013) 073081.
17. K-Y Kim, Origin of the Abraham spin angular momentum of surface modes, Opt. Lett. 39 (2014) 682–684.
18. K.-Y. Kim, A. X. Wang, Spin angular momentum of surface modes from the perspective of optical power flow, Opt. Lett. 40 (2015) 2929–2932.
19. K.-Y. Kim, A. X. Wang, Relation of the angular momentum of surface modes to the position of their power-flow center, Opt. Express 22 (2014) 30184–30190.
20. T. G. Philbin, Electromagnetic energy-momentum in dispersive media, Phys. Rev. A 83 (2012) 013823; T. G. Philbin, Phys. Rev. A 85 (2012) 059902(E) (Erratum).
21. T. G. Philbin, O. Allanson, Optical angular momentum in dispersive media, Phys. Rev. A 86 (2012) 055802.
22. K. Y. Bliokh, A. Y. Bekshaev, F. Nori, Optical momentum and angular momentum in dispersive media: From the Abraham-Minkowski debate to unusual properties of surface plasmon-polaritons, New J. Phys. 19 (2017) 123014.
23. K. Y. Bliokh, A. Y. Bekshaev, F. Nori, Optical momentum, spin, and angular momentum in dispersive media, Phys. Rev. Lett. 119 (2017) 073901.
24. K. Y. Bliokh, A. Y. Bekshaev, Spin and momentum of light fields in dispersive inhomogeneous media with application to the surface plasmon-polariton wave, Ukr. J. Phys. Opt. 19 (2018) 33–48.
25. A. Y. Bekshaev, M. S. Soskin, Transverse energy flows in vectorial fields of paraxial beams with singularities, Opt. Commun. 271 (2007) 332–348.
26. M. V. Berry, Optical currents, J. Opt. A: Pure Appl. Opt. 11 (2009) 094001.
27. A. Y. Bekshaev, K. Y. Bliokh, M. S. Soskin, Internal flows and energy circulation in light beams, J. Opt. 13 (2011) 053001.
28. J. D. Jackson, Classical Electrodynamics, Wiley, New York, 1999.
29. L. D. Landau, E. M. Lifshitz, L. P. Pitaevskii, Electrodynamics of Continuous Media, Pergamon, New York, 1984.
30. A. Aiello, P. Banzer, M. Neugebauer, G. Leuchs, From transverse angular momentum to photonic wheels, Nature Photon. 9 (2015) 789–795.
31. A. Aiello, P. Banzer, The ubiquitous photonic wheel, J. Opt. 18 (2016) 085605.
32. R. Ruppin, Electromagnetic energy density in a dispersive and absorptive material, Phys. Lett. A 299 (2002) 309–312.
33. S. Huard, C. Imbert, Measurement of exchanged momentum during interaction between surface-wave and moving atom, Opt. Commun. 24 (1978) 185–189.

34. T. Matsudo, Y. Takahara, H. Hori, T. Sakurai, Pseudomomentum transfer from evanescent waves to atoms measured by saturated absorption spectroscopy, Opt. Commun. 145, (1998) 64–68.
35. K. Bliokh, F. Nori, Transverse and longitudinal angular momenta of light, Physics Reports 592 (2015) 1–38.
36. Y. G. Song, S. Chang, J. H. Jo, Optically induced rotation of combined Mie particles within an evanescent field of a Gaussian beam, Jpn. J. Appl. Phys. 38 (1999) L380–L383.
37. Y. G. Song, B. M. Han, S. Chang, Force of surface plasmon-coupled evanescent fields on Mie particles, Opt. Commun. 198 (2001) 7–19.
38. L. Wei, M. F. Picardi, J. J. Kingsley-Smith, A. V. Zayats, F. J. Rodríguez-Fortuño, Directional scattering from particles under evanescent wave illumination: the role of reactive power, arXiv:1803.04821 (2018).
39. A. Y. Bekshaev, K. Y. Bliokh, F. Nori, Transverse spin and momentum in two-wave interference, Phys. Rev. X 5 (2015) 011039.
40. M. Antognozzi *et al*, Direct measurements of the extraordinary optical momentum and transverse spin-dependent force using a nano-cantilever, Nat. Phys. 12 (2016) 731–735.
41. R. Hertel, Theory of the inverse Faraday effect in metals, J. Magn. Magn. Mat. 303 (2006) L1–L4.
42. E. M. Purcell, Electricity and Magnetism (Berkeley Physics Course, Vol. 2), McGraw-Hill, New York, 1985.
43. A. Herczyński, Bound charges and currents, Am. J. Phys. 81 (2013) 202–205.
44. A. D. Rakic, A. B. Djurisic, J. M. Elazar, M. L. Majewski, Optical properties of metallic films for vertical-cavity optoelectronic devices, Appl. Opt. 37 (1998) 5271–5283.
45. L. I. Schiff, Quantum mechanics, McGrow-Hill, New York, 1968.
46. M. G. Silveirinha, N. Engheta, Transformation electronics: Tailoring the effective mass of electrons, Phys. Rev. B 86 (2012) 161104.
47. Y. C. Jun et al, Active tuning of mid-infrared metamaterials by electrical control of carrier densities, Opt. Express 20 (2012) 1903–1911.
48. H.-I. Lee, J. Mok, Orbital and spin parts of energy currents for electromagnetic waves through spatially inhomogeneous media, J. Mod. Opt. 65 (2018) 1053–1062.
49. A. Y. Bekshaev, Dynamical characteristics of an electromagnetic field under conditions of total reflection, J. Opt. 20 (2018) 045604.